\documentclass[%
reprint,
superscriptaddress,
amsmath,amssymb,
aps,
pre,
floatfix,
]{revtex4-2}

\usepackage[american]{babel}
\usepackage{graphicx}
\usepackage{dcolumn}
\usepackage{bm}
\usepackage[version=4]{mhchem} 
\usepackage{siunitx} 
\DeclareSIUnit{\molar}{M}
\usepackage{xspace} 
 
\newcommand*{\dd}{\mathrm{d}}
\newcommand*{\logTen}{\log_{10}}
\newcommand*{\caIon}{\ce{Ca^{2+}}}
\newcommand*{\caConcentration}{\ce{[Ca^{2+}]}\xspace}
\newcommand*{\caFifty}{\ce{[Ca^{2+}]_{50}} }
\newcommand*{\pca}{\text{pCa}}
\newcommand*{\concentration}{c}

\newcommand*{\Tiso}{T_0}
\newcommand*{\TisoMax}{T_{0}^{\text{max}}}
\newcommand*{\Fiso}{F_{0}}

\newcommand*{\hCoeff}{n_{\text{H}}}
\newcommand*{\avgSpin}{\langle s\rangle}
\providecommand{\avg}[1]{\langle #1\rangle}
\newcommand*{\kbt}{k_{\text{B}}T}
\newcommand*{\konoff}{k_{\pm}}
\newcommand*{\kon}{k_{+}}
\newcommand*{\koff}{k_{-}}
\providecommand{\state}[2]{\bigl(\begin{smallmatrix}#1\\#2\end{smallmatrix}\bigr)}
\providecommand{\rate}[3]{k_{#1,#2}^{#3}}

\begin{document}

\preprint{AIP/123-QED}

\title{Ising models of cooperativity in muscle contraction}

\author{Elaheh Saadat}
\affiliation{SISSA, via Bonomea 265, Sezione di Trieste, 34136 Trieste, Italy}

\author{Matthieu Caruel}
\affiliation{Université Paris-Est Créteil, Université Gustave Eiffel, CNRS, UMR 8208, MSME, F-94010 Créteil, France}
\email{matthieu.caruel@u-pec.fr}

\author{Stefano Gherardini}
\affiliation{Istituto Nazionale di Ottica del Consiglio Nazionale delle Ricerche (CNR-INO), Largo Enrico Fermi 6, I-50125 Firenze, Italy}
\email{stefano.gherardini@ino.cnr.it}

\author{Ilaria Morotti}
\affiliation{Department of Biology, University of Florence, I-50019 Sesto Fiorentino, Italy}
\affiliation{PhysioLab, University of Florence, I-50019 Sesto Fiorentino, Italy}

\author{Matteo Marcello}
\affiliation{Department of Biology, University of Florence, I-50019 Sesto Fiorentino, Italy}
\affiliation{PhysioLab, University of Florence, I-50019 Sesto Fiorentino, Italy}

\author{Marco Caremani}
\affiliation{Department of Biology, University of Florence, I-50019 Sesto Fiorentino, Italy}
\affiliation{PhysioLab, University of Florence, I-50019 Sesto Fiorentino, Italy}

\author{Marco Linari}
\affiliation{Department of Biology, University of Florence, I-50019 Sesto Fiorentino, Italy}
\affiliation{PhysioLab, University of Florence, I-50019 Sesto Fiorentino, Italy}

\author{Ivan Latella}
\affiliation{Departament de Física de la Matèria Condensada, Universitat de Barcelona, Martí i Franquès~1, 08028 Barcelona, Spain}
\affiliation{Institut de Nanociència i Nanotecnologia de la Universitat de Barcelona (IN2UB), Diagonal 645, 08028 Barcelona, Spain}
\email{ilatella@ub.edu}

\author{Stefano Ruffo}
\affiliation{SISSA, via Bonomea 265, Sezione di Trieste, 34136 Trieste, Italy}
\affiliation{Istituto dei Sistemi Complessi, Consiglio Nazionale delle Ricerche, Via Madonna del Piano 10, and INFN,
Sezione di Firenze, 50019 Sesto Fiorentino, Italy}
\email{ruffo@sissa.it}

\begin{abstract}
    Regulation of contraction in striated muscle is controlled by a dual mechanism involving both thin filaments containing actin and thick filaments containing myosin.
    The thin filament is activated by calcium ions binding to troponin, leading to tropomyosin azimuthal displacement which allows the activation of a regulatory unit (composed of one troponin, one tropomyosin and seven actin monomers) that exposes the actin sites for interaction with the myosin motors.
    Motor attachment to actin contributes to spreading activation within and beyond a regulatory unit along the thin filament through a cooperative mechanism.
    We introduce a one-dimensional Ising model to elucidate the mechanism of cooperativity in thin filament activation in relation to the force generated by the attached myosin motor.
    The model characterizes thin filament activation and cooperativity using only two parameters: one related to calcium concentration and the other to the force exerted by the attached myosin motor, which is modulated by temperature.
    At any force, the model is able to determine the extent of actin-myosin interactions on a correlation length ranging from two to seven actin monomers in addition to the seven actin monomers of the regulatory unit.
    Our theoretical predictions are successfully tested on experimental data, and our tests also include the condition of hindered filament activation by the use of the specific drug Omecamtiv Mecarbil (OM).
    According to our model, the effect of OM results in an anticooperativity mechanism accounting for the  experimental data.
\end{abstract}

\keywords{Muscle activation, Ising model, Troponin, Omecamtiv Mecarbil}

\maketitle

\section{Introduction}
\label{sec:introduction}

\begin{figure}
    \centering
    \includegraphics{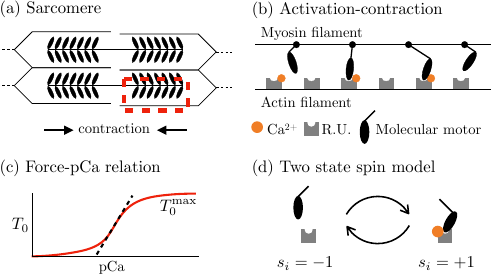}
    \caption{
        (a) Schematic of a muscle sarcomere. The red rectangle surrounds interacting myosin and actin filaments. (b) Activation-contraction mechanism: \caIon ions bind to regulatory units (RUs) to allow molecular motor attachment and force generation; we stress that the RUs are only pictorially represented.
        (c) Typical force-pCa relation (red continuous line). The maximum slope of the relation (black dashed line), associated to the Hill coefficient \(\hCoeff \), quantifies the degree of cooperativity. (d) Two-state Ising model used in this paper.
    }
    \label{fig-context}
\end{figure}

In the striated muscle cell, steady force and shortening are due to two bipolar arrays of the motor protein myosin II emerging from the backbone of the thick filament.
The arrays of myosin motors pull the thin actin-containing filaments toward the center of the sarcomere, the \(\qty{\sim 2}{\micro\meter} \)-long structural unit of striated muscle, during cyclical ATP-driven interactions~\cite{huxley-1973}, as shown in Figs.~\ref{fig-context}(a) and \ref{fig-context}(b).
In the resting muscle, contraction is prevented by the regulatory protein tropomyosin (Tm) in the thin filament that masks the actin sites for myosin binding~\cite{gordon-2000}. Contraction is also prevented by the autoinhibition that operates in the myosin motors in the OFF state, in which the motors lie on the surface of the thick filament folded on their tail toward the center of the sarcomere unable to bind actin and hydrolyze ATP~\cite{woodhead-2005,stewart-2010}.
Upon stimulation, \caIon is released from the sarcoplasmic reticulum and binds to the regulatory protein troponin C (TnC) of the troponin complex (Tn) on the thin filament. This allows for an azimuthal displacement of Tm that exposes the actin sites for the interactions with the myosin motors~\cite{huxley-1973a}. During loaded contractions, the stress generated by a few constitutively ON motors in the thick filament would allow the OFF motors to be switched ON~\cite{linari-2015,reconditi-2017,piazzesi-2018,brunello-2020}.
Thick filament mechanosensing is likely mediated by titin that, upon stimulation, has recently been shown to become an efficient mechanical coupler at physiological sarcomere length~\cite{squarci-2023,morotti-2024}.

Along the thin filament, \caIon binding to TnC allows the activation of regulatory units (RUs). For each strand of the double-stranded helix, an RU is composed of seven actin monomers, one Tn complex and one Tm. Activation can spread beyond an RU for the head-tail interactions between consecutive Tm's, suggesting a cooperative mechanism of thin filament activation.
Experimental evidence shows that strong attachment of the myosin motor to actin is responsible for full actin site availability~\cite{weber-1973,swartz-1992,vibert-1997,craig-2001}. This supports a two-step steric-blocking model of thin filament activation~\cite{mckillop-1993}.
During the first step, \caIon binding to TnC is responsible for partial displacement of Tm (blocked-to-closed transition) that, within one RU, allows myosin to establish weak interactions with the actin.
The second step is triggered by the weak-to-strong transition of  myosin-actin interaction that further promotes Tm displacement (closed-to-open transition), which may provide the cooperative mechanism for spreading thin filament activation beyond the RU~\cite{desai-2015,longyear-2017,caremani-2022}.

The modulation of the muscle contraction by calcium can be characterized by the relation between the \caIon concentration (\caConcentration) and the steady force \(\Tiso \) generated in isometric conditions by demembranated muscle fibers at a given sarcomere length~\cite{moisescu-1976}.
This dependence of the force on the \caConcentration \ is represented by the so-called {\it force-pCa relation,} where \(\pca = -\logTen(\caConcentration) \), see Fig.~\ref{fig-context}(c).

As in various biological regulatory processes, the relation between the activation input (pCa) and its output (\(\Tiso \)) is highly non-linear, showing a sharp sigmoidal transition between all-or-none states.
Denoting \(\TisoMax \), the isometric force measured at saturating \(\caConcentration \), the normalized force \(f = \Tiso/\TisoMax \)
is often conventionally associated with the Hill function~\cite{hill-1910}, i.e.,
\begin{equation}
    \label{eq:hill_function}
    f = \frac{1}{1+c^{-\hCoeff}}\,,
\end{equation}
where \(c = \caConcentration/\caFifty\), such that \(\Tiso = \TisoMax/2 \) when \( c=1 \).
The exponent \(\hCoeff \), called the Hill coefficient, represents the cooperativity degree of the activation process, which can be associated with the slope of the force-pCa relation around \(\caConcentration=\caFifty \), as shown in Fig.~\ref{fig-context}(c).
A high cooperativity translates into a sharp increase in the tension upon a small change in the calcium concentration.

Recently, Caremani \emph{et al.}~\cite{caremani-2022} reported that \(\hCoeff \) increases proportionally with the force \(\Fiso \) exerted by a myosin motor attached to actin; such a force is modulated by temperature.
This observation suggests that the force produced by the attached motors may enhance the activation of neighboring regulatory units without the help of extra calcium binding.
This mechanistic explanation of the observed increase in cooperativity with \(\Fiso \) is supported by measurement of the force-pCa relation in the presence of the positive inotrope Omecamtiv Mecarbil (OM)~\cite{malik-2011,nagy-2015}. OM is a small molecule that suppresses the myosin working stroke and prolongs the lifetime of the myosin-actin attachment at physiological ATP concentrations and at therapeutic OM concentrations~\cite{nagy-2015,woody-2018,governali-2020a}. In the presence of OM, \(\hCoeff \) shows a shallower slope of the force-pCa relation, which may be attributed to OM-bound motors failing to generate force~\cite{woody-2018,governali-2020a}.

The objective of this paper is to provide a physical model that is able to validate this interpretation of the experimental results, by also providing quantitative predictions that may stimulate further experiments. For this purpose, we use a statistical mechanical approach based on the classical one-dimensional Ising model.
Our approach aims to evaluate the effect of temperature and of the presence of OM on the parameters used in the one-dimensional Ising model to quantify cooperativity in isometric steady state conditions.
Because of its simplicity, this model is insufficient to describe
experimental data underpinning the current understanding of the molecular processes behind muscle contraction. More comprehensive models can be found in Refs.~\cite{hill-1974,hill-1976a,mijailovich-2012,smith-2008,smith-2008a,caremani-2013,mijailovich-2021a,stojanovic-2019,liu-2023,regazzoni-2020,filipovic-2022}.
In the current literature, an Ising-like model of thin filament activation was formulated by Rice \emph{et al.}~\cite{rice-2003}, who considered a model of spins structured in two layers, where each unit of the model represents a RU. The first layer of spins identifies the presence of absence of \caIon on the RU, while the second layer describes whether the RU is occupied by a force-generating myosin motor.

In the present work, we show that the force-pCa relations measured experimentally by Caremani \emph{et al.}~\cite{caremani-2022} can be well-reproduced already with a single-layer Ising spin model, see Fig.~\ref{fig-context}(d).
We relate the change in the degree of cooperativity, observed at different temperatures and in the presence/absence of OM, to the change in the spread of thin filament activation between neighboring RUs. In particular, a large cooperativity suggests a molecular process by which the number of RU activated by a \caIon ion can be larger than one through head-tail interactions between neighboring Tn-Tm complexes, as previously suggested by Regnier \emph{et al.}~\cite{regnier-2002}.

Moreover, we establish a bijective relationship between the Hill coefficient and the coupling constant of the single-spin Ising model, which can be interpreted as an effective measure of cooperativity at the molecular scale. We show that a modulation of this coupling constant accounts for the dependence of the Hill coefficient on the motor force \(\Fiso \) observed experimentally in \cite{caremani-2022}.
Furthermore, the computation of the correlation length of the Ising model indicates that the range of these interactions increases with \(\Fiso \) and is of the order of one to two regulatory units, which supports the mechanistic model proposed in \cite{caremani-2022}. We also obtain that larger temperatures correlate with stronger nearest-neighbor interactions of the model.

Finally, we quantitatively compare our results with the former work by Rice \emph{et al.}~\cite{rice-2003}.
While their model was expected to take into account more faithfully the physiological activation mechanism, we show that it is over-determined, at least in the context set by the available data on force-pCa relations obtained on skeletal muscle.
Moreover, we demonstrate that the Rice \emph{et al.} model cannot be calibrated to reproduce simultaneously, with the same non-degenerate set of parameters, the behaviors of the force-pCa relations observed at high- and low-calcium concentrations.

\section{Two-state Ising model}
\label{sec:two-state-model}

We consider a one-dimensional chain of \(N \) force-generating units with nearest-neighbor interactions.
Each unit represents an actin-myosin complex that can be in two states: \emph{non-force generating} in which no \caIon\ is bound, and the motor is detached, or \emph{force generating} in which \caIon is bound and a myosin motor is attached and produces force.
The state of unit \(i\in\{1,\dots,N\} \) is characterized by a discrete variable (spin) \(s_i\in \{-1,+1\} \), where \(s_i=+1 \) (respectively \(-1 \)) corresponds to the force generating (respectively non-force-generating) state, see Figs.~\ref{fig-context}(b) and \ref{fig-context}(d).

In the following, we consider that the parameters entering our model are dimensionless, with all energy quantities being rescaled by the thermal energy \(\kbt \).
As suggested by the experimental study conducted in \cite{caremani-2022}, we introduce a dimensionless coupling constant \(J \) characterizing the strength of nearest-neighbor interactions. We denote by \(h \) the dimensionless external field, accounting for the concentration of \caIon, and write the Hamiltonian of the system as
\begin{equation}
    H_1 = - \sum_{i=1}^N (J\,s_{i}s_{i+1} + h s_{i}),
\end{equation}
with periodic boundary conditions.
Following the approach used by \cite{zwanzig} and \cite{LiviBook}, the spin model can be associated with a two-state Markov process with transition rates \( \konoff^i\). This mapping reads as
\begin{equation*}
    \ce{$s$_{i} = -1 <=>[$\kon^{i}$][$\koff^{i}$] $s$_{i} = +1}, \quad \forall i\in\{1,\dots,N\},
\end{equation*}
where the transition rates \( \konoff^{i}\) linked to the \(i \)-\emph{th} unit verify the detailed balance condition
\begin{eqnarray}\label{eq:detailed_balance_hamiltonian}
    \frac{\kon^{i}}{\koff^{i}} &=& \exp\Big\{\!-\big[ H_1(s_i=+1) - H_1(s_i=-1) \big]\Big\}=\nonumber \\
    &=& \exp\Big\{ 2\big[ J\left(s_{i+1}+s_{i-1}\right) + h\big] \Big\}\,.
\end{eqnarray}
To compare our model predictions with the experimental results, we associate (i) the external field \(h \) to the normalized intracellular calcium concentration \(\concentration\) and (ii) the coupling \(J \) to the Hill coefficient \(\hCoeff \) using the identities
\begin{equation}
    \label{eq:relation_h_c_nh_j}
    h = \frac{1}{2}\log(\concentration),\quad\quad J = \frac{1}{2}\log(\hCoeff)\,.
\end{equation}
With these notations, the detailed balance condition \eqref{eq:detailed_balance_hamiltonian} becomes
\begin{equation}
    \label{eq:detailed-balance-two-state}
    \frac{\kon^{i}}{\koff^{i}} = \concentration \, \hCoeff^{(s_{i+1}+s_{i-1})}\,.
\end{equation}

\subsection{Force-pCa relation}

Quantities of interest in the one-dimensional Ising model can be obtained using the standard transfer matrix method.\footnote{Interested readers can refer, for instance, to pp. 204--210 of Baxter's book \cite{Baxter}, or to \cite{Nishimori,pelishke}} For instance, the magnetization per spin is given by
\begin{equation}
    \label{eq:average_spin}
    \avgSpin = \frac{\sinh(h)}{\left[\sinh^{2}(h)+e^{-4J}\right]^{1/2}}\,,
\end{equation}
which ranges in the interval $[-1,1]$.
Considering that units in state \(+1 \) can generate force, the normalized force \(f=\Tiso/\TisoMax\) is equal to the average fraction of permissive units: \(f = (1+\avgSpin)/2 \). Taking into account Eqs.~\eqref{eq:relation_h_c_nh_j} and \eqref{eq:average_spin}, this normalized force can be rewritten as
\begin{equation}\label{eq:average_force_single_spin}
    f = \frac{1}{2}\left\{ 1 + \frac{c-1}{\big[ (c-1)^{2} + 4\,\hCoeff^{-2}c \big]^{1/2}}\right\}.
\end{equation}
As expected, the above expression is compatible with the tension reaching half of its maximal value at \(c=1 \), i.e., \(\caConcentration=\caFifty \), which justifies the expression of \(c \) introduced in \eqref{eq:relation_h_c_nh_j}.

In both the conventional Hill model \eqref{eq:hill_function} and the two-state model \eqref{eq:average_force_single_spin},
the Hill coefficient can be recovered using
\begin{equation}
    \label{eq:hcoeff_definition}
    \hCoeff = \left.\frac{\dd}{\dd \log c}\log\left(
    \frac{f}{1-f}
    \right)\right|_{c=1}.
\end{equation}
Furthermore, in the limits \(c\to 0 \) and \(c\to\infty \), the function \(\logTen\left[f/(1-f)\right]\) depends linearly on \(\pca = -\logTen(\caConcentration)\) with slope 1 in both cases.
Hence, the model can be fully calibrated using the measured value of \(\logTen[f/(1-f)] \) in the region near \(c=1 \), where this function can be reasonably approximated by a straight line parametrized by the reference concentration \(\caFifty \) and the slope \(\hCoeff \).

\begin{figure}
    \centering
    \includegraphics{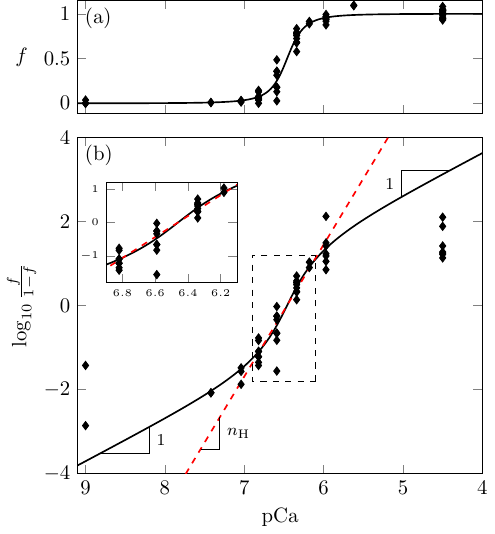}
    \caption{
        pCa-dependence of the normalized force \( f \) in panel (a) and \( \logTen\left(\frac{f}{1-f}\right) \) in panel (b) (black continuous line) obtained after calibration of the model to experimental data (symbols) measured at \SI{25}{\degreeCelsius}.
        The experimental data are normalized by the average tension measured at $\text{pCa}=\num{4.5}$.
        The insert in (b) shows the data chosen to fit the parameter \(\hCoeff \) using linearization at \(c=1 \), so that the results are not affected by the deviation of the data from the prediction at the extremes of the interval. The best fit [red dashed line in (b)] is achieved setting
        \(\caFifty = \qty[parse-numbers = false]{10^{-6.5}}{\molar} \) and \(\hCoeff=3.14 \).
        For one fiber, the value of the force at $\text{pCa}=\num{9}$ was measured twice. For that fiber, the mean of the two values was subtracted from the force to all data points generating a negative and a positive value of the same amplitude at pCa 9.
    }
    \label{fig-force-pca}
\end{figure}

The calibration is illustrated in Fig.~\ref{fig-force-pca}, where the force-pCa relation predicted by our model is compared to experimental data obtained from demembranated fibers of rabbit soleus muscle at \SI{25}{\degreeCelsius}, see \cite{caremani-2022,force_pca_data}.
To normalize our experimental data, we used the average force measured at $\pca=4.5$ (saturating concentration) as the reference tension and \(\caFifty = 10^{-6.5}\unit{\molar} \) as our reference concentration.
The value of \(\hCoeff \) was then estimated from the subset of data shown in the inset of Fig.~\ref{fig-force-pca}(b), using both a custom implementation of the minimum least squares error methods and the built-in non-linear fitting method from Mathematica \cite{nonlinearmodelfit}, both giving similar results.
We obtained \(\hCoeff=3.14 \) or, equivalently, \(J= \num{0.57} \) according to Eqs.~\eqref{eq:relation_h_c_nh_j}.

The calibration procedure was repeated on similar data from \cite{caremani-2022} at \qtylist{12;17;25;35}{\degreeCelsius}, and also in the presence of \SI{1}{\micro\molar} OM.
It is worth stressing that, for each set of force-pCa curves, the free parameters whose value we fitted are the coupling constant $J$ and the \( \caFifty \). Interestingly, the values of \caFifty resulting from the fits do not vary much with temperature.

\begin{figure}
    \centering
    \includegraphics{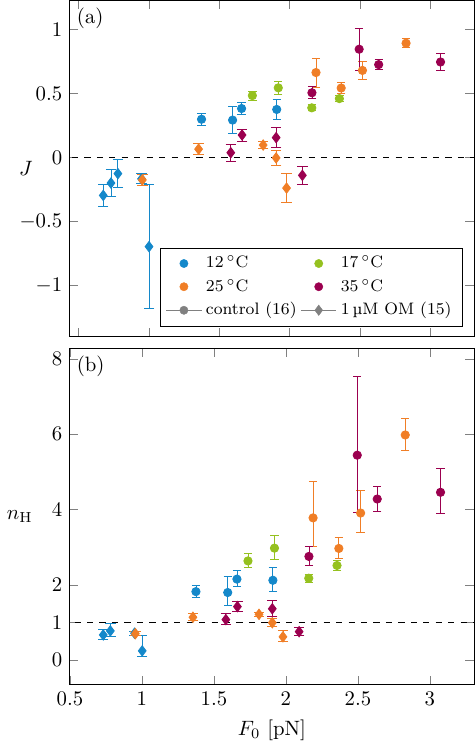}
    \caption{\label{fig-hillcoefficient} Dependence of the coupling constant  \(J \) (a) and of the Hill coefficient \(\hCoeff \) (b) on the motor force modulated by temperature without OM (circles) and in the presence of \SI{1}{\micro\molar} OM (diamonds).
        Each symbol represents a measurement from a single muscle fiber \cite{force_pca_data}.
        The results are shown as a function of the force per motor \(F_0\) measured for different fibers. Experimental data are taken from \cite{caremani-2022}. Error bars indicate the error in the calibration. In total, 15 and 16 measurements were performed with and without OM, respectively. The number of fibers used for each temperature is indicated in parentheses in the legend.
    }
\end{figure}

The fitted values of $J$ and of the Hill coefficients are represented in Fig.~\ref{fig-hillcoefficient} against the force per motor \(\Fiso\).
The value of \(\Fiso \) is determined from experiments using the expression \( \Fiso = \kappa_0 x_0 \), where \( \kappa_0 \) and \( x_0 \) denote the stiffness and elongation of a single motor, respectively.
Both \( \kappa_0 \) and \( x_0 \) can be estimated by means of specific mechanical protocols by which the contribution of myosin motors and myofilaments to the half-sarcomere compliance during active isometric contraction can be identified \cite{caremani-2022}.
An important experimental finding is that \(\Fiso \) is independent of \caConcentration, while \(\Tiso \) increases with \caConcentration in proportion to the number of attached motors \(N_0 \): \(\Tiso = N_{0}\Fiso \)\cite{linari-2007,caremani-2022}.
Hence, at each temperature, \(\Fiso \) can be considered as a parameter modulating the cooperativity of the system i.e., how fast \(N_0 \) increases with \caConcentration.

For fibers without OM (control), the Hill coefficient almost triples as the temperature ranges from \SI{17}{\degreeCelsius} to \SI{35}{\degreeCelsius}, signaling a significant increase in cooperativity, see Fig.~\ref{fig-hillcoefficient}.
The obtained values of $J$ and \(\hCoeff\) are similar between \SI{25}{\degreeCelsius} and \SI{35}{\degreeCelsius}.
Conversely, in the presence of OM (diamonds), the Hill coefficient takes values smaller than 1 for the temperatures \qtylist{12;25;35}{\degreeCelsius}.
Hence, the presence of OM seems to suppress the cooperative activation. In fact, values of $\hCoeff$ smaller than $1$ implies that $J < 0$. Therefore, the interaction term in the one-spin Ising model becomes anti-ferromagnetic. It entails that the presence of a force-generating unit disfavors a nearby one, so that the system becomes anticooperative in this regime.

These findings support the description of the cooperative mechanism put forward in \cite{caremani-2022}, according to which the force generated by an attached motor modulates the probability of activating neighboring RUs, thereby affecting the Hill coefficient.

\subsection{Correlation between neighboring units}

\begin{figure}
    \centering
    \includegraphics{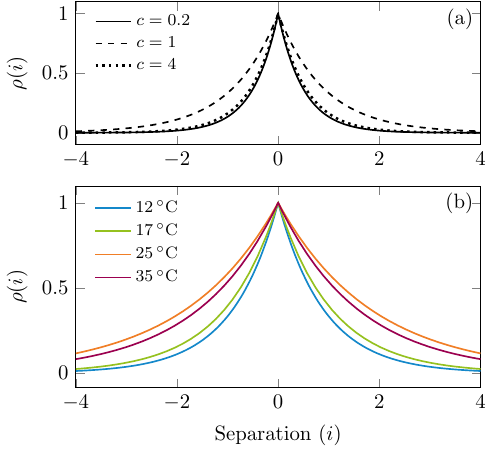}
    \caption{
        \label{fig-correlation} (a) Correlation function at different calcium concentrations obtained from the model fitted to the experimental data at \SI{12}{\degreeCelsius}.
        (b) Correlation function at \(c=1 \) for the temperatures \qtylist{12;17;25;35}{\degreeCelsius} presented in Fig.~\ref{fig-hillcoefficient} in control conditions.
    }
\end{figure}

To further characterize the cooperative mechanism, we introduce the  correlation function
\begin{equation}\label{eq:rho_i}
    \rho(i):=\frac{\langle s_{0} s_{i}\rangle - \langle s\rangle^2 }{\langle s_{0}^{2}\rangle - \langle s\rangle^2},
\end{equation}
which quantifies the tendency for a remote site \(i \) to be in the same configuration as an arbitrary reference site~0.
For the sake of simplicity, we discuss only the case of $J \ge 0$ (i.e., $\hCoeff \ge 1$) for which we have \(\rho(i) \) = \((\lambda_-/\lambda_+)^i \),
where \(\lambda_\pm = e^{ J}\big\{\cosh(h) \pm \big[\sinh^{2}(h) + e^{-4 J}\big]^{1/2}\big\} \) are the eigenvalues of the transfer matrix~\cite{Baxter}.

The spreading of the correlation function is maximum at \(c=1 \), see Fig.~\ref{fig-correlation}(a), coherently with the fact that this concentration also corresponds to the highest slope of the force-pCa curve, see Fig.~\ref{fig-force-pca}(a).
Furthermore, the correlation length increases with the force per motor, as expected from the increase in the cooperativity, see Fig.~\ref{fig-correlation}(b). At low forces (temperature \qty{12}{\degreeCelsius}), the influence of the single unit is negligibly small beyond the fourth neighbor, while it largely exceeds this range at higher forces (temperatures \(>\qty{25}{\degreeCelsius}\)).

\begin{figure}
    \centering
    \includegraphics{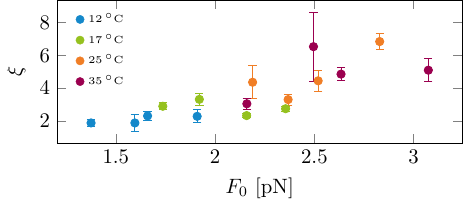}
    \caption{
        Correlation length $\xi$ as a function of the force per motor \(F_0\) modulated by temperature.
    }
    \label{fig-correlation_length}
\end{figure}

The dependence of the cooperative mechanism on the force per motor can be characterized by the correlation length~\cite{Baxter,Nishimori} \( \xi = \left[ \log{(\lambda_+/\lambda_-)} \right]^{-1} \), which reads as
\begin{equation}
    \xi = \left[ \log{ \left(\frac{\hCoeff +1}{\hCoeff -1}\right) } \right]^{-1}
    \label{eq:corrlength}
\end{equation}
when \(c=1 \). As for the Hill coefficient (see Fig.~\ref{fig-hillcoefficient}), we obtain an increase of the correlation length $\xi$ with the increase of the force per motor \(\Fiso \), from \(\xi\approx 2 \) at \qty{12}{\degreeCelsius} to \(\xi\approx 6 \) at \(\qtylist{25;35}{\degreeCelsius} \) in control conditions, see Fig.~\ref{fig-correlation_length}.

Overall, the above analysis of the correlations between neighboring single (two-state) spin models in the presence or absence of OM tends to support the hypothesis that spreading of the activation signal to neighboring RUs is due to the increase of motor force at submaximal calcium concentration.

\section{Comparison with a four-state model}

As mentioned in the Introduction, a description for the activation of units generating force was proposed by Rice \emph{et al.}~\cite{rice-2003}, introducing a model of spins structured in two layers. Below, after briefly reviewing their approach, we reformulate it in terms of a generalized external field that allows for a simpler interpretation of the output force and a comparison with our model.

\begin{figure}[htbp]
    \centering    \includegraphics{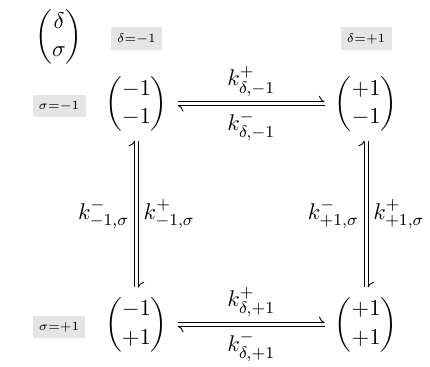}
    \caption{Four-state model proposed by Rice \emph{et al.}~\cite{rice-2003}. The system consists of a 1D chain of \(N \) units, each being described with two spins \(\sigma =\pm 1\) and \(\delta=\pm 1 \) generating four possible states. The transitions between the different states are characterized by the set of rate constants \(\rate{\delta}{\sigma}{\pm} \).}
    \label{fig:four-state-model}
\end{figure}

\subsection{Original formulation}

The system in~\cite{rice-2003} is represented as a chain of $N$ units, each being characterized by two spins: \(\sigma =\pm 1\) and \(\delta=\pm 1 \).
The spin \(\delta \) takes the value \(+1 \) when a \caIon\ is bound, and the value \(-1 \) if not.
The spin \(\sigma \), instead, takes the value \(+1 \) if a myosin head can attach and generate force (i.e., the unit is permissive), and the value \(-1 \) if the myosin head cannot attach (i.e., the unit is nonpermissive).
Together, \(\delta \) and \(\sigma \) define four states \( \state{\delta}{\sigma}\) schematized in Fig.~\ref{fig:four-state-model}.
The configuration of the whole system is then characterized by the set of spins of all units \(\left\{\state{\delta_{i}}{\sigma_{i}},i\in\{1,\dots,N\}\right\} \), which, for an actin filament constituted of 26 regulatory units, represents \(4^{26}\) distinct configurations.

The dimensionless Hamiltonian of this two-spin system can be written as~\cite{rice-2003}
\begin{equation}
    \label{eq:hamiltonian_rice}
    H_2 = - \sum_{i=1}^{N} \left( J \sigma_{i} \sigma_{i+1} +  K \sigma_{i} \delta_{i} +  h_{\delta} \delta_{i} + h_{\sigma} \sigma_{i}  \right),
\end{equation}
where \(h_{\delta}\) and \(h_{\sigma}\) play the role of external fields associated with the spins \(\delta_i\) and \(\sigma_i\), respectively.
The parameter \(K\) is a coupling for spins of different species on site $i$, and \(J\) accounts for nearest-neighbor interactions for spins \(\sigma_i\).
The associated jump process is characterized by eight transition rates \(\rate{\delta}{\sigma}{\pm} \), see Fig.~\ref{fig:four-state-model}.
For instance, the rate \(\rate{-1}{\sigma}{+}\) (respectively, \(\rate{-1}{\sigma}{-} \)) is associated with the transition from the state \(\state{-1}{-1} \) (respectively, \(\state{-1}{+1} \)) to state \(\state{-1}{+1} \) (respectively, \(\state{-1}{-1} \)), see Fig.~\ref{fig:four-state-model}.

As shown in Ref.~\cite{rice-2003}, the equilibrium response of the system can be described in terms of the parameters $C$, $Q$, $\mu$, and $\gamma$ defined by
\begin{align*}
    \frac{1}{2}\log{(C)}   & =  h_{\delta} -  K, & \frac{1}{2} \log{(Q)}     & = h_{\sigma} + K, \\
    \frac{1}{4}\log{(\mu)} & = K,                & \frac{1}{2}\log{(\gamma)} & = J.
\end{align*}
The average normalized force produced by the system is given by the fraction of units in the permissive state: \( f = (1+\avg{\sigma})/2 \), which thus combines the populations in states \(\state{-1}{+1} \) and \(\state{+1}{+1} \). In particular, taking the large \(N \) limit, the average normalized force takes the form~\cite{rice-2003}
\begin{equation}
    \label{eq:force_rice}
    f = \frac{1}{2} \left\{ 1 + \frac{A C + B}{\big[ (A C + B)^2 + 4 D  (C+1) (\mu C + 1) \big]^{1/2}} \right\},
\end{equation}
where
\begin{align*}
    A & = \mu^{1/4} \left(Q^{1/2} - Q^{-1/2}\right),      \\
    B & = \mu^{-3/4} \left(Q^{1/2} - \mu Q^{-1/2}\right), \\
    D & = \gamma^{-2} \mu^{-1/2}.
\end{align*}

\subsection{Reformulation}

Here, we choose a different parametrization of the previous four-state model by introducing the following quantities:
\begin{subequations}
    \label{eq:parameters-four-state}
    \begin{align}
        \frac{1}{2}\log(c) & = h_{\sigma} + h_{\delta}, \quad\quad\quad
        \frac{1}{2}\log(p) = h_{\sigma} - K,                            \\
        \frac{1}{2}\log(q) & = h_{\delta} - K, \quad\quad\quad
        \frac{1}{2} \log{(\hCoeff)} = J\,.
    \end{align}
\end{subequations}
With this new set of parameters, the force \eqref{eq:force_rice} can now be written (see Appendix \ref{sec:effective_field}) as
\begin{equation}
    \label{eq:force_rice_redressed}
    f = \frac{1}{2}\left\{ 1 +
    \frac{x-1}{
        \big[ (x-1)^{2} + 4\hCoeff^{-2}x \big]^{1/2}
    }
    \right\},
\end{equation}
which is analogous to \eqref{eq:average_force_single_spin} using the equivalent external field
\begin{equation}
    x = \frac{c+p}{q+1}\,.
\end{equation}
This result shows that, like in the two-state Ising model, the force is determined by only two independent parameters, namely \(x \) and \(\hCoeff \). Hence, the data of the force-pCa relation alone cannot fully determine the four-state model.

\subsection{Detailed balance conditions and calcium dependence}

Using the mapping \eqref{eq:parameters-four-state}, we can write the detailed balance conditions associated with the different transition rates represented in Fig.~\ref{fig:four-state-model} (see Appendix \ref{sec:detailed_balance}) as
\begin{subequations}
    \label{eq:detailed-balance-four-state}
    \begin{align}
        \frac{\rate{\delta}{-1}{+}}{\rate{\delta}{-1}{-}} & = q,
                                                          & \frac{\rate{\delta}{+1}{+}}{\rate{\delta}{+1}{-}} & = \frac{c}{p},                                     \\
        \frac{\rate{-1}{\sigma}{+}}{\rate{-1}{\sigma}{-}} & =p\,\hCoeff^{(\sigma_{i-1}+\sigma_{i+1})},
                                                          & \frac{\rate{+1}{\sigma}{+}}{\rate{+1}{\sigma}{-}} & =\frac{c}{q}\hCoeff^{(\sigma_{i-1}+\sigma_{i+1})}.
    \end{align}
\end{subequations}
To compare the model \eqref{eq:hamiltonian_rice} with experimental data, Rice \emph{et al.}~consider that the concentration in calcium ions affects only the kinetics of the transition \(\delta_i = -1\to\delta_i =+1 \), which means in our case that \(p \) is independent on \caConcentration.
Assuming first-order kinetics, Rice \emph{et al.}~considered \( q = \frac{\rate{\delta}{-1}{+}}{\rate{\delta}{-1}{-}} = \frac{\caConcentration}{K_{d}} \), where \(K_{d} \) is the calcium dissociation constant in the nonpermissive states. Similarly, \( \frac{c}{p} = \frac{\rate{\delta}{+1}{+}}{\rate{\delta}{+1}{-}} = \frac{\caConcentration}{K_{d}'} \) (if $p\neq 0$), where \(K_{d}' \) is the calcium dissociation constant in the permissive states.
These hypotheses reveal the dependence of our effective field \(x \) on $[{\rm Ca}^{2+}]$
(provided $p\neq 0$):
\begin{equation}
    x = \frac{K_{d}}{p K_{d}'} \frac{\left( \caConcentration + K_{d}' \right)}{\left(\caConcentration + K_{d}\right)}\,.
\end{equation}

We first notice that in the particular case where \( K_{d}=K_{d}' \), we have \(x=1/p\), which is independent on \caConcentration.
Second, when \(\caConcentration\to\infty \), \(x\to\frac{K_{d}}{pK_{d}'} \), which implies that the system does not necessarily reach the maximum tension \(f=1 \) even at large \(\caConcentration \).
Third, when \(\caConcentration\to 0 \), \(x\to\frac{1}{p} \), which implies that the system might generate force even without calcium.
Interestingly, the system may reach \(f=1 \) at saturating calcium if \(p\to 0 \), which can be achieved if \(h_{\sigma}\to-\infty \).
Conversely, the system can reach \(f=0 \) at low calcium concentration if \(p\to\infty\), obtained if \(h_{\sigma}\to +\infty \). Otherwise, for other values of $h_{\sigma}$, the correct limits at low and high calcium concentrations are not correctly reproduced by the model.

\subsection{Reduced two-state models}

We now study how the generated force depends on the coupling \(K \) between the two spins of the Rice \emph{et al.}~model. First, we consider complete uncoupling by setting \(K=0 \). In such a case, we have \(x = p = \exp[2h_{\sigma}] \), which means that the effective field is the one associated with the spin \(\sigma \), see \eqref{eq:hamiltonian_rice}. In particular, in this limit, the force becomes independent of \caConcentration.

Second, we now consider the strong coupling limit \(K\to\infty \), which implies \(p\to 0 \) and \(q\to 0 \) so that \(x = c  = \exp[2(h_{\sigma}+h_{\delta})]\).
Recalling the detailed balance relations \eqref{eq:detailed-balance-four-state}, the limits \(p\to0 \) and \(q\to 0 \) imply that the heterogeneous states \(\state{-1}{+1} \) and \(\state{+1}{-1} \) (bottom left and top right corners in Fig.~\ref{fig:four-state-model}) will not be populated, leaving only the homogeneous states \(\state{-1}{-1} \) and \(\state{+1}{+1} \).
This result can also be inferred from the Hamiltonian \eqref{eq:hamiltonian_rice} since high values of \(K \) will favor the homogeneous states.
Therefore, the strong coupling limit leads to a single spin model with \(s_{i} = \state{-1}{-1}\) or \(s_{i} =\state{+1}{+1}\), whose response does depend on calcium concentration, and which is then fully analogous to the one-spin (two-state) model presented in Sec.~\ref{sec:two-state-model}.

To support this statement, we compute the equilibrium constant of the two reaction paths \([\state{-1}{-1}\to\state{+1}{-1}\to\state{+1}{+1}] \) and \([\state{-1}{-1}\to\state{-1}{+1}\to\state{+1}{+1}] \). For both paths we find:
\begin{equation}\label{eq:overline_K}
    \frac{[\state{+1}{+1}]}{[\state{-1}{-1}]} = \frac{\rate{\delta}{-1}{+}}{\rate{\delta}{-1}{-}}\frac{\rate{+1}{\sigma}{+}}{\rate{+1}{\sigma}{-}} = \frac{\rate{-1}{\sigma}{+}}{\rate{-1}{\sigma}{-}}\frac{\rate{\delta}{+1}{+}}{\rate{\delta}{+1}{-}} = c\,\hCoeff^{(s_{i-1}+s_{i+1})}
\end{equation}
that corresponds to the detailed balance \eqref{eq:detailed-balance-two-state} under the combined field \(h_{\sigma}+h_{\delta} \) and the identification of the coupled spins \( \state{\delta_{i}}{\sigma_{i}}\) with \( s_i \).
Notice that Eq.~\eqref{eq:overline_K} is also valid  when \(K \) is finite, meaning that the only field that controls the ratio of populations in states \(\state{-1}{-1} \) and \(\state{+1}{+1} \) is \(h_{\sigma} + h_{\delta} \), independently of the coupling between the spins.

\section{Discussion}

\subsection{Alternative models}

This work presents a model of the thin filament activation for muscle contraction based on a single spin (two-state) Ising model where a unit can be either ``nonforce-generating'' or ``force-generating.''
The model incorporates two key parameters: a control parameter \(h \) linked to the experimental calcium concentration, and a free parameter \(J \) that determines the energy gain resulting from having two consecutive units in the same state. Therefore, \(J \) quantifies cooperativity.
These phenomenological parameters are not directly related to the actual molecular mechanisms at work during activation; they are introduced to effectively reproduce the behavior observed in experimental data, as customary for such systems.

In terms of complexity, our two-state Ising model has direct correspondence with the classical Hill model, which also relies on two parameters: the reference concentration \caFifty and the Hill coefficient \(\hCoeff \). The advantage of the Ising  model lies in its ability to provide a mechanistic interpretation of the Hill coefficient in terms of nearest-neighbor interactions. It also allows for the computation of correlation functions and correlation lengths, which are not accessible through the Hill model.

After normalization of \(h \) using the reference calcium concentration \caFifty that is directly obtained from experiments, the behavior of our model is fully determined by the sole coupling constant \(J \). We have thus shown that the resulting one-parameter model is sufficient to accurately reproduce the experimental force-pCa curves.

More detailed representations of thin filament activation have been proposed in the literature, often implementing the so-called ``blocked-closed-open'' paradigm introduced by McKillop and Geeves~\cite{mckillop-1993}.
The corresponding models are usually built upon the Huxley-Hill-type framework for modeling unregulated actin-myosin interactions~\cite{hill-1974}.
The Huxley-Hill-type models capture time-dependent behaviors and can be calibrated using a large variety of experimental observations ranging from single molecule experiments to \emph{in vitro} motility assays and fiber mechanics.
The cooperative filament activation can then be accounted for by \emph{ad hoc} variations of the attachment rate of unbound motors with the distance from already attached motors~\cite{longyear-2017,woody-2018,walcott-2014,debold-2013}.
Once the parameters of the underlying actin-myosin interaction model are calibrated using data obtained at saturating \caConcentration, the modulation of actin-myosin interactions is then adjusted to match data obtained by varying \caConcentration. This modulation can effectively rely on a few additional parameters.

In all cases, these formulations involve defining rate functions that
may be associated with a virtually infinite number of parameters. These rate functions can eventually be parametrized by a few scalars that nevertheless necessitate more experimental data than the sole force-pCa curves considered in the present work to be specifically identified.

Furthermore, in the framework of Ising-like models, accounting for the blocked-closed-open paradigm requires at least two spin variables, as proposed by Rice \emph{et al.}~\cite{rice-2003}, but we demonstrate here that such four-state model can be mapped into a simpler two-state version for the fitting of force-pCa relation.

In summary, the model used in this paper can be viewed as a drastically reduced version of these more comprehensive approaches, where we assume that both activation states and actin-myosin interaction states can be lumped in two effective states, and that the dynamics of transitions between these states can be neglected in favor of an equilibrium description for the steady state force-pCa curves.

\subsection{Cooperativity}

Despite its simplicity, our approach successfully captures the characteristic features of the force-pCa curve, including the upper and lower limits, and its sigmoidal shape.
The model leverages nearest-neighbor interactions within the Ising framework to elucidate how cooperativity operates at the filament level.
The coupling coefficient $J$ quantifies the energy required, in units of $\kbt$, to match the observed effect of temperature changes on the degree of cooperativity illustrated by variation in both the Hill coefficient and the correlation length between neighboring regulatory units.
The increased correlation length can be interpreted as a stronger head-tail molecular interaction between consecutive tropomyosin proteins along the actin filament \cite{risi-2024,risi-2023,yamada-2020}.

\subsection{Effect of omecamtiv mecarbil}

Our analysis reveals distinct behaviors for fibers in control conditions and fibers treated with OM.
Notably, the cooperativity parameter becomes negative in the presence of OM, suggesting that the drug not only disrupts cooperativity but can also introduce anticooperativity within the thin filament.
In the present context, anticooperativity is defined by a Hill coefficient \(\hCoeff < 1\) following the terminology used for protein-ligand binding systems, or equivalently a negative coupling constant \(J<0 \) as in antiferromagnetic systems.

This finding aligns with the experimental observations at \qty{12}{\degreeCelsius} from Refs.~\cite{caremani-2022,governali-2020a}, reporting values of \(\hCoeff = \num[uncertainty-mode = separate]{0.79(9)} \) and \(\hCoeff = \num[uncertainty-mode = separate]{0.84(15)} \), respectively.
At room temperature, the effect of OM is less pronounced with observed values down to \(\hCoeff \num{\approx 1}\) \cite{nagy-2015,caremani-2022,governali-2020a}, though a value slightly below \num{1} (\num[uncertainty-mode = separate]{0.97(7)}) is reported for rat cardiomyocytes
in Ref.~\cite{nagy-2015}.

The effect of OM on force generation by the actin-myosin system has been studied at the molecular level by Woody et al. \cite{woody-2018} based on the model developed in \cite{longyear-2017,debold-2013,walcott-2014}.
The experimental data obtained by Woody \emph{et al.} on single molecules can then be reproduced by modulating the kinetics of the actin-myosin interaction, particularly by reducing the detachment rate of OM-bound myosin heads and the power-stroke size.
They show that this modulation can account for the effect of OM on calcium sensitivity, as materialized by horizontal shifts of the force-pCa curve.
However, the effect of OM on cooperativity, as measured by the Hill coefficient is not explicitly addressed.

The results obtained by Caremani \emph{et al.} \cite{caremani-2022} and Governali \emph{et al.} \cite{governali-2020a} suggest that variation in the Hill coefficient with OM cannot be explained solely by the reduction of the force per motor induced by OM binding, and therefore that OM would not affect only single motor properties as postulated by Woody \emph{et al.}
Hence, their result supports the hypothesis that OM indirectly modulates the interaction between neighboring units, which is precisely what we quantify through variation of the coupling constant with OM.

\subsection{Limitations}

A potential limitation of this work is that it is based on an approach that is valid under thermodynamic equilibrium conditions.
Physiologically, muscle consumes energy even during isometric contraction, where no mechanical work is performed, which implies that the system breaks the detailed balance condition \cite{julicher-1997,wang-2002}. Nevertheless, in this work we implicitly assume that effective free energy landscapes can be defined, an approximation that is also guided by the fact that no external action is exerted on the system once the steady state is reached.

The next step in our research involves establishing a quantitative relationship between thin filament cooperativity and the availability of the myosin motor from the nearby thick filaments with the aim to achieve a deeper understanding of the interplay between thin filament cooperativity and thick filament regulation~\cite{brunello-2023,morotti-2024,ma-2024}.
The predictions of the model should also be tested using experimental data from different muscle types and species, and in different experimental conditions to assess its generality.

\begin{acknowledgments}
    We thank Vincenzo Lombardi and Massimo Reconditi for continuous discussion and critical reading of the manuscript. I.L. acknowledges financial support from the Spanish Government through Grants No. PID2021-126570NB-I00 and No. PID2024-156516NB-I00 financed by MICIU/AEI/10.13039/501100011033 and FEDER/UE. S.G., S.R., and M.L. acknowledge financial support from the European Union — NextGeneration EU within PRIN 2022, PNRR, Project No. P2022XPT32 Regulation of striated muscle: a research bridging single molecule to organ (ResTriMus) (CUP B53D23033290001).
\end{acknowledgments}

\section*{Data availability}
The data that support the findings of this article are openly available \cite{force_pca_data}.

\onecolumngrid

\appendix

\section{Effective field}
\label{sec:effective_field}

We now discuss the mathematical steps to obtain Eq.~\eqref{eq:force_rice_redressed}. For this purpose, it is convenient to express the normalized force in terms of the parameters of the Hamiltonian~\eqref{eq:hamiltonian_rice} of the two-spin system. The direct application of the transfer matrix method entails the following expression:
\begin{equation*}
    \label{eq:force_rice_hamiltonian_parameter}
    f = \frac{1}{2} \left( 1 + \frac{e^{h_{\sigma}+2h
            _{\delta}} + e^{h_{\sigma}-2K} - e^{2h_{\delta}-h_{\sigma}-2K}  -  e^{-h_{\sigma}}}{\Big( \left(e^{h_{\sigma}+2h
            _{\delta}} + e^{h_{\sigma}-2K} - e^{2h_{\delta}-h_{\sigma}-2K}  -  e^{-h_{\sigma}}\right)^2  + 4 \, e^{-4J}  \left(e^{4h_{\delta}-2K} + e^{2h_{\delta}-4K} + e^{2 h_{\delta}} + e^{-2K}\right)\Big)^{1/2}} \right).
\end{equation*}
Thus, factorizing the numerator and the denominator by \( \exp(-h_{\sigma}) \) and using the definitions \eqref{eq:parameters-four-state} leads us straightforwardly to \eqref{eq:force_rice_redressed} with the effective field \(x=\frac{c+p}{q+1} \).

\section{Detailed balance relation}
\label{sec:detailed_balance}

In this Appendix, we report the rationale that entails the detailed balance conditions associated with the transition rates represented in Fig.~\ref{fig:four-state-model}.
Denoting \( \sigma_{i \pm 1} = \sigma_{i-1} + \sigma_{i+1} \), equations \eqref{eq:detailed-balance-four-state} are obtained by explicitly writing the identities
\begin{equation*}
    \begin{split}
        \frac{\rate{\delta}{j}{+}}{\rate{\delta}{j}{-}} & = \exp\big( -H_2(\delta_i=+1,\sigma_i=j) + H_2(\delta_i=-1,\sigma_i=j) \big)\,,  \\
        \frac{\rate{j}{\sigma}{+}}{\rate{j}{\sigma}{-}} & = \exp\big( - H_2(\delta_i=j,\sigma_i=+1) + H_2(\delta_i=j,\sigma_i=-1) \big)\,,
    \end{split}
\end{equation*}
with $j\in\{+1,-1\}$, namely
\begin{equation*}
    \begin{split}
        \frac{\rate{\delta}{-1}{+}}{\rate{\delta}{-1}{-}} & = \exp\big( -(K(-1)(-1) - h_{\delta}) + (K(+1)(-1) + h_{\delta}) \big) = \exp\big( 2(h_{\delta} - K) \big) = q\,,    \\
        \frac{\rate{\delta}{+1}{+}}{\rate{\delta}{+1}{-}} & = \exp\big( -(K(-1)(+1) - h_{\delta}) + (K(+1)(+1) + h_{\delta}) \big) = \exp\big( 2(h_{\delta} + K) \big)
        = \frac{\exp\big( 2(h_{\delta} + h_{\sigma}) \big) }{\exp\big( 2(h_{\sigma}-K) \big)} = \frac{c}{p}\,,                                                                   \\
        \frac{\rate{-1}{\sigma}{+}}{\rate{-1}{\sigma}{-}} & = \exp\big( -J(\sigma_{i\pm1})(-1) -K(-1)(-1) - h_{\sigma} (-1) + J(\sigma_{i\pm1}) + K(-1)(+1) + h_{\sigma} \big) = \\&= \exp\big( 2J\sigma_{i\pm1}\big) \exp\big(2 h_{\sigma} - 2K \big) = p\,\hCoeff^{\sigma_{i\pm 1}}\,,\\
        \frac{\rate{+1}{\sigma}{+}}{\rate{+1}{\sigma}{-}} & =
        \exp\big( -J(\sigma_{i\pm1})(+1) -K(+1)(-1) - h_{\sigma} (-1) + J(\sigma_{i\pm1}) + K(+1)(+1) + h_{\sigma} \big)=                                                        \\
                                                          & = \exp\big( 2J\sigma_{i\pm1} \big) \exp\big(2 h_{\sigma} + 2K \big)
        = \hCoeff^{\sigma_{i\pm 1}} \frac{\exp\big( 2(h_{\delta} + h_{\sigma})\big)}{\exp\big( 2(h_{\delta}-K)\big)} = \frac{c}{q}\hCoeff^{\sigma_{i\pm 1}}.
    \end{split}
\end{equation*}

\twocolumngrid

\providecommand{\noopsort}[1]{}\providecommand{\singleletter}[1]{#1}%

\end{document}